\documentclass[a4paper,10pt]{article}

\usepackage{subfig}
\usepackage{graphicx}
\usepackage{cite}
\usepackage{epsfig,amsmath,amssymb,verbatim,mathrsfs,array,layout,textcomp,amssymb,latexsym}

\newcommand{\beq}{\begin{eqnarray}}
\newcommand{\eeq}{\end{eqnarray}}
\newcommand{\be}{\begin{eqnarray}}
\newcommand{\ee}{\end{eqnarray}}
\newcommand{\nn}{\nonumber}
\def\beqa{\begin{eqnarray}}
\def\eeqa{\end{eqnarray}}
\def\bea{\begin{eqnarray}}
\def\eea{\end{eqnarray}}
\newcommand{\no}{\nonumber}
\newcommand{\bv}{\left(\begin{array}{c}}
\newcommand{\ev}{\end{array}\right)}
\newcommand{\bmtwo}{\left(\begin{array}{cc}}
\newcommand{\bmthree}{\left(\begin{array}{ccc}}
\newcommand{\emn}{\end{array}\right)}
\newcommand{\bmtwoc}{\left\{\begin{array}{cc}}
\newcommand{\bmthreec}{\left\{\begin{array}{ccc}}
\newcommand{\emnc}{\end{array}\right\}}
\newcommand{\ba}{\begin{array}}
\newcommand{\ea}{\end{array}}

\def\lsim{\mathrel{\rlap{\lower4pt\hbox{\hskip1pt$\sim$}}
     \raise1pt\hbox{$<$}}}         
\def\gsim{\mathrel{\rlap{\lower4pt\hbox{\hskip1pt$\sim$}}
     \raise1pt\hbox{$>$}}}         

\begin{document}

\begin{titlepage}

\vskip1.5cm
\begin{center}
{\Large \bf Probing scalar dark matter oscillations with neutrino oscillations}
\end{center}
\vskip0.2cm

\begin{center}
Marta Losada$^1$, Yosef Nir$^2$,  Gilad Perez$^2$ and Yogev Shpilman$^2$\\
\end{center}
\vskip 8pt

\begin{center}
{ \it $^1$New York University Abu Dhabi, PO Box 129188, Saadiyat Island, Abu Dhabi, United Arab Emirates\\
$^2$Department of Particle Physics and Astrophysics,\\
Weizmann Institute of Science, Rehovot 7610001, Israel} \vspace*{0.3cm}

{\tt   marta.losada@nyu.edu, yosef.nir,gilad.perez,yogev.shpilman@weizmann.ac.il}
\end{center}

\vglue 0.3truecm

\begin{abstract}
  \vskip 3pt \noindent
If ultra-light dark matter (ULDM) exists and couples to neutrinos, it can be discovered via time-periodic variations in the neutrino mass and mixing parameters. We analyze the current bounds on such a scenario and establish the sensitivity expected for both time-averaged and time-resolved modulations in future neutrino oscillation experiments. We place a special emphasis in our analysis on time modulations of the CP violating mixing phase. We illustrate with a toy model the case where the leading modulation effect can be CP violating while the  effect on CP conserving parameters is suppressed. We show a unique imprint that a time averaged CP violating modulation of ULDM can leave in neutrino oscillations, while direct CP asymmetries vanish.
\end{abstract}

\end{titlepage}

\section{Introduction}
Dark matter (DM) is responsible for about a quarter of the energy density of the Universe. Nevertheless, the fundamental nature of DM is still not established.
Furthermore, several solid theoretical reasons imply that some of the fundamental constants in nature are in fact dynamical and can be effectively described as expectation values of scalar fields (see~\cite{Svrcek:2006yi,Uzan:2010pm} and Refs. therein for relevant discussions). If these scalar fields are sufficiently light, they may change in time even today, leading to time-varying `constants' that can be searched for at the precision frontier (see {\it e.g.}~\cite{Safronova:2017xyt} for a recent review).

A particularly concrete and phenomenologically attractive possibility is that the light scalar field forms a viable ultra-light DM (ULDM) candidate~\cite{Arvanitaki:2014faa,Graham:2015ifn}, in which case the field amplitude (and as a result some of the fundamental constants) would oscillate at the scalar Compton frequency, $\omega_{\rm{C}}=m_{\phi}$, where $m_{\phi}$ is the scalar mass. Theoretically, constructing a natural model of a scalar ULDM is rather challenging. However, two concrete proposals have been put forward, one where the DM mass is protected by an approximate scale-invariance symmetry~\cite{Arvanitaki:2014faa}, and a second one where it is protected by an approximate shift-symmetry that is broken, together with CP~\cite{Flacke:2016szy}, by two sequestered sectors~\cite{Banerjee:2018xmn} (inspired by the relaxion paradigm~\cite{Graham:2015cka}). The two models are qualitatively very different, yet, in both frameworks the DM field couples to the SM  either due to the fact that its coupling breaks scale-invariance~(see for instance~\cite{Goldberger:2007zk}) or via mixing with the Higgs field~\cite{Flacke:2016szy}. If the ULDM mass is extremely light, $m_\phi\lesssim 10^{-21}\,$eV, then it may be probed model-independently via its gravitational interactions \cite{Bozek:2014uqa,Armengaud:2017nkf,Irsic:2017yje,Zhang:2017chj,Kobayashi:2017jcf,Bar:2018acw,Bar:2019bqz}. Currently, however, neither observations nor theoretical arguments constrain the DM mass~(see for instance \cite{Bertone:2018xtm}), and thus broad-band based searches, which require couplings between the DM field and the SM ones, are motivated. 
In the specific models of scalar ULDM mentioned above, the coupling between the DM and the SM is present and calculable. Yet this coupling is expected to be small because of both direct bounds from equivalence-principle (EP) tests and naturalness considerations (see for instance~\cite{Banerjee:2020kww} for a recent discussion). A plethora of direct detection experiments are being proposed and developed to search for ULDM depending on its interaction with SM particles.

In this work we are interested in exploring the consequences of the possible interactions of ULDM with neutrinos \cite{Krnjaic:2017zlz,Brdar:2017kbt,Capozzi:2018bps,Berlin:2016woy,Dev:2020kgz}. In specific models there could be a situation where the dark matter would couple to the neutrino sector more strongly than what is naively expected based on the neutrino's dilatonic charge or the $U(1)$ charges~\cite{Huber:2002gp,Perez:2008ee,Davidi:2018sii}, which makes the corresponding signal a bit more pronounced, in addition to the fact that the EP-based bounds tend to be weaker. (We comment on this issue towards the end of this paper.) Regardless of these theoretical considerations, in this work we take a more phenomenological approach and just assume that the DM couples to the neutrino sector and study the resulting sensitivity to such coupling in various experiments, see~\cite{Krnjaic:2017zlz,Brdar:2017kbt,Capozzi:2018bps,Berlin:2016woy,Dev:2020kgz} for earlier works following a similar approach.


In our analysis below we shall focus on a DM candidate with sub-eV mass, such that we can treat it as a classical bosonic field that oscillates with time,
\beq
\phi\approx\phi_0\sin(m_\phi t)\,,
\eeq
with
\beq
\phi_0\simeq\frac{\sqrt{2\rho_\phi^{\oplus}}}{m_\phi}\sim2\times10^{10}\ {\rm GeV}\left(\frac{10^{-22}\ {\rm eV}}{m_\phi}\right),
\eeq
$\rho_\phi^{\oplus}$ corresponds to the ULDM density on the surface of the Earth, where gravity-based measurements yield a relatively weak bound on it (see for instance~\cite{Hogan:1988mp,  Banerjee:2019epw, Anderson:2020rdk} for relevant discussions) and  also the corresponding DM field amplitude is subject to stochastic fluctuations around its commonly assumed preferred value~\cite{Foster:2017hbq, Centers:2019dyn}.
Nevertheless, for concreteness we assumed $\rho_\phi^{\oplus}=\rho_\phi^\odot\sim0.3\ {\rm GeV}/{\rm cm}^3$, and $\rho_\phi^\odot$ is the DM local-galactic density with coherent time that is equal to $1/m_\phi\beta_\odot^2$, with $\beta_\odot\sim220\,$km/sec\,.
The oscillation period is given by
\beq\label{eq:tauphi}
\tau_\phi=\frac{2\pi}{m_\phi}\approx1.3\ {\rm year}
\times\left(\frac{10^{-22}\ {\rm eV}}{m_\phi}\right)\,.
\eeq
In this work we consider neutrino oscillation experiments with relevant time scales that range between ${\cal O}(10^{-3}\ {\rm second})$ and ${\cal O}(1\ {\rm year})$. This translates into a corresponding range in $m_\phi$ of
\beq
10^{-22}\ {\rm eV}\lsim m_\phi\lsim 10^{-12}\ {\rm eV}.
\eeq
The upper bound guarantees that the ULDM field remains approximately constant along the propagation distance typical to long baseline neutrino oscillation experiments, of ${\cal O}(10^3\ {\rm km})$ and the lower bound correspond to the fuzzy DM bound~\cite{Hu:2000ke}.

The plan of this paper goes as follows. In Section \ref{sec:uldmnu} we present our theoretical framework and demonstrate the effects of interest in a two neutrino generation model. In Section \ref{sec:experiments} we list the neutrino oscillation experiments of relevance to our study and their relevant features. In Section \ref{sec:timeaveraged} we study the effects of fast modulations of the $\phi$-field and the current bounds. In Section \ref{sec:cpv} we introduce a toy model that leads to time modulations that are larger in CP violating parameters than in the CP conserving ones, and obtain the current bounds and expected future sensitivities to such a scenario. In Section \ref{sec:timeresolved} we study the effects of slow modulations of the $\phi$-field, the current bounds and the expected future sensitivities.  The statistical method that we apply to this scenario is presented in Appendix \ref{app:slow}. Section \ref{sec:pheno} discusses the phenomenological constraints from naturalness and scalar DM-Higgs mixing for this model and briefly comments on the cosmological bound. We summarize our results in Section \ref{sec:summary}.

\section{ULDM effects on neutrino oscillations}
\label{sec:uldmnu}
We consider extending the SM with a gauge-singlet scalar field $\phi$. We are particularly interested in its effect on neutrino physics. We thus consider the following dimension-five and dimension-six terms in the Lagrangian:
\beq\label{eq:defzy}
{\cal L}_{z,y}=\frac{z_{\alpha\beta}}{\Lambda}\left(L^{\alpha}\right)^{T}L^{\beta}HH+\frac{y_{\alpha\beta}}{\Lambda^{2}}\phi\left(L^{\alpha}\right)^{T}L^{\beta}HH,
\eeq
where $L(1,2)_{-1/2}$ is the lepton doublet field, $H(1,2)_{+1/2}$ is the Higgs field, $z$ and $y$ are $3\times3$ symmetric matrices of dimensionless couplings, and $\Lambda$ is a scale of new physics. Replacing the Higgs field with its VEV, $\langle H\rangle=(0\ v/\sqrt2)^T$, we obtain the following mass and Yukawa terms for neutrinos:
\beq
{\cal L}_{m_{\nu}}=\frac{z_{\alpha\beta}v^{2}}{2\Lambda}\left(\nu^{\alpha}\right)^{T}\nu^{\beta}+\frac{y_{\alpha\beta}v^{2}}{\Lambda^{2}}\phi\left(\nu^{\alpha}\right)^{T}\nu^{\beta}.
\eeq
Thus, the neutrino mass matrix and the effective Yukawa matrix are given by
\beqa\label{eq:defmyhat}
m_\nu&=&zv^2/\Lambda,\\
\hat y&=&yv^2/\Lambda^2.\no
\eeqa
Treating $\phi$ as a classical field, it modifies the neutrino mass matrix:
\beq
\hat m_\nu=m_\nu+\hat y\phi.
\eeq

In order to gain some intuition for the implication of the $\phi\nu\nu$ couplings, we consider the two-neutrino case, and work in the basis where $m_\nu$ is diagonal. We assume that all entries of the matrix $\hat y\phi$ are much smaller than $\Delta m\equiv m_2-m_1$. The effective mass-squared difference is given by
\beqa\label{eq:deltahat}
\Delta\hat m^2&=&\Delta m^2+2(m_2\hat y_{22}-m_1\hat y_{11})\phi+{\cal O}\left(\hat y^2\phi^2\right)\no\\
&\equiv&\Delta m^2\left[1+2\eta_{\Delta}\sin(m_\phi t)\right].
\eeqa
The effective mixing angle is given by
\beqa\label{eq:thetahat}
\hat\theta&=&\theta+\frac{\hat y_{12}\phi}{\Delta m}+{\cal O}\left(\frac{\hat y^2\phi^2}{m^2}\right)\no\\
&\equiv&\theta+\eta_\theta\sin(m_\phi t).
\eeqa
Note that both $\eta_{\Delta}$ and $\eta_\theta$ are ${\cal O}\left(\frac{y}{z}\frac{\phi_0}{\Lambda}\right)$. Thus, for example, with $\Lambda\sim10^{12}\ {\rm GeV}$, $z\sim10^{-3}$ and $y\sim0.1$, it is possible to obtain $m_\nu\sim0.05$ eV simultaneously with $\eta={\cal O}(1)$.

We now proceed with the evaluation of the transition probability $P_{\mu e}$ to ${\cal O}(\hat y\phi)$. We define
\beq
x_E\equiv\frac{\Delta m^2 L}{4E},\ \ \
\hat x_E\equiv\frac{\Delta\hat m^2 L}{4E}.
\eeq
The  $\nu_{\mu} \rightarrow \nu_e$ appearance  oscillation probability is given by
\beq\label{eq:pmuehat}
P_{\mu e}=\sin^22\hat\theta\sin^2\hat x_E.
\eeq
We make a further approximation, by assuming that $\delta x_E\equiv\hat x_E-x_E\ll1$.
Before we proceed with the analysis, let us estimate $x_E$ quantitatively. One can write
\beq
x_E=1.27\frac{\Delta m^2}{{\rm eV}^2}\frac{L}{\rm km}\frac{{\rm GeV}}{E}.
\eeq
For DUNE, the relevant parameters are $E\sim{\rm GeV}$ and $L\sim1300$ km. Thus,
\beq
x_E^{\rm DUNE}\sim\frac{\Delta m^2}{6\times10^{-4}\ {\rm eV}^2}.
\eeq
For $\Delta m^2_{21}\sim7.4\times10^{-5}\ {\rm eV}^2$, indeed $x^{\rm DUNE}_{E21}\sim0.1$. With our assumption that $\delta x_E\ll x_E$, it is safe to take $\sin\delta x_E\sim\delta x_E$. For $\Delta m^2_{31}\sim2.5\times10^{-3}\ {\rm eV}^2$, $x^{\rm DUNE}_{E31}\sim4$. Here it is therefore questionable whether we can take $\sin\delta x_E\sim\delta x_E$. Nevertheless, we provisionally continue to make this approximation.

Inserting Eqs. (\ref{eq:deltahat}) and (\ref{eq:thetahat}) into the expression (\ref{eq:pmuehat}) gives
\beq\label{eq:hatpmue}
P_{\mu e}=\sin^22\theta\sin^2 x_E
+\left(2\eta_\theta\sin4\theta\sin^2x_E
+2\eta_\Delta x_E\sin^22\theta\sin2x_E\right)\sin(m_\phi t).
\eeq

Let us now separate the cases of $\theta$-modification and $\Delta m^2$ modification.
\begin{itemize}
\item Taking $\hat y_{11}=\hat y_{22}=0$, then to linear order in $\hat y_{12}\phi/m$ only the mixing angle is affected:
\beq\label{eq:hatpmuetheta}
P_{\mu e}=\sin^2 x_E\sin^2\left\{2\left[\theta+\eta_\theta\sin(m_\phi t)\right]\right\}.
\eeq
\item Taking $\hat y_{12}=0$, then to linear order in $\hat y_{22}\phi/m$ only the mass-squared difference is affected:
\beq\label{eq:hatpmuedelta}
P_{\mu e}=\sin^22\theta\sin^2\left\{x_E[1+2\eta_\Delta\sin(m_\phi t)]\right\}.
\eeq
\end{itemize}

\section{Neutrino oscillation experiments}
\label{sec:experiments}
When considering neutrino oscillation experiments, in addition to $\tau_\phi$, given in Eq. (\ref{eq:tauphi}), there are several experiment-specific time scales
that play a role:
\begin{itemize}
\item[$\tau_d$] - the source-to-detector distance.
\item[$\tau_r$] - the time-resolution of the detector. We take it to be the time needed in order to accumulate a statistically significant number of $\sim 10/${events}. This time scale may vary at a given experiment when one considers different energy bins of different event rates. We therefore provide a rough estimate for $\tau_r$ in various experiments, and the exact value needs to be calculated separately for the probed bin.
\item[$\tau_e$] - the running time of the experiment.
\end{itemize}

For $\tau_\phi<\tau_d$, the effective neutrino mass and mixing change along the propagation from source to detector. The analysis of this complicated case is beyond the scope of this work. For $\tau_\phi>\tau_e$, the experiment is insensitive to the effects of the $\phi$ field. Thus, we will be interested in the range
\beq\label{eq:tauphirange}
\tau_d \lsim \tau_\phi \lsim \tau_e,
\eeq

For $\tau_{\phi}$ within the range of Eq. (\ref{eq:tauphirange}), one can  distinguish two regimes:
\begin{itemize}
\item $\tau_\phi<\tau_r$: ``Fast modulations" or, more precisely, ``time-averaged modulations". The time dependence of the variation of the neutrino parameters cannot be resolved, and only an averaged effect can be observed.
\item $\tau_\phi>\tau_r$: ``Slow modulations" or, more precisely, ``time-resolved modulations". The time dependence of the variation of the neutrino parameters can be resolved.
\end{itemize}

\subsection{Accelerator neutrinos: DUNE and HK}
The next generation long baseline neutrino experiments DUNE and Hyper-Kamiokande (HK), will use accelerator neutrinos. The transition probability $P_{\mu e}^0$ (the super-index $0$ means ``in the absence of the $\phi$-field'') can be approximated as \cite{Nunokawa:2007qh}
\beqa\label{eq:pmuedune}
P_{\mu e}^0&=&\sin^2\theta_{23}\sin^22\theta_{13}\sin^2x_{E31}\no\\
&+&\sin2\theta_{23}\sin2\theta_{13}\sin2\theta_{12}x_{E21}\sin x_{E31}\cos(x_{E31}+\delta)\no\\
&+&\cos^2\theta_{23}\sin^22\theta_{12}x_{E21}^2.
\eeqa

At DUNE, the expected unoscillated event rate\footnote{By "unoscillated event rate" we mean the expected rate if the transition/survival probability were 1.} is 6000/year \cite{Diwan:2013eha} and $L=1300$ km. The relevant time scales are the following:
\beqa
\tau_d&\approx&4.3\times10^{-3}\ {\rm sec},\no\\
\tau_r&\approx&16\ {\rm weeks}\sim1.2\times10^6\ {\rm sec},\no\\
\tau_e&\approx&7\ {\rm years}\sim2\times10^8\ {\rm sec}.
\eeqa
At HK, the expected unoscillated event rate\footnote{We approximated the unoscillated event rate using the expected oscillated event rate from \cite{Kearns:2013lea} and normalizing with $P_{\mu e}$.}  is 20580/year and $L=295$ km. The relevant time scales are the following:
\beqa
\tau_d&\approx&9.8\times10^{-4}\ {\rm sec},\no\\
\tau_r&\approx&1\ {\rm day}\sim8.6\times10^4\ {\rm sec},\no\\
\tau_e&\approx&2.5\ {\rm years}\sim9\times10^7\ {\rm sec}.
\eeqa
The actual planned running time of HK is 10 years which are split in a $1:3$ ratio between neutrino and antineutrino modes such that the total number of events would be approximately the same in both of them. We consider here the more intense neutrino mode.
Eq.~(\ref{eq:tauphi}) implies that, for DM-$\phi$, $\tau_\phi\not\gg\tau_e$.

\subsection{Reactor neutrinos: Daya Bay, KamLAND, and JUNO}
Daya Bay, KamLAND, and JUNO measure the flux of reactor neutrinos.  In this case the disappearance transition probability $P_{\bar e\bar e}^0$ is given by
\beq\label{eq:peedayabay}
P_{\bar e\bar e}^0=1-\cos^4\theta_{13}\sin^22\theta_{12}\sin^2x_{E21}-\sin^22\theta_{13}\sin^2x_{E31}.
\eeq
The expected unoscillated event rate is 800/day in Daya Bay \cite{An:2016ses}, 2/day in KamLAND \cite{Suekane:2004ny}, and 84/day in JUNO \cite{Giaz:2018gdd}.\\
The relevant time scales at Daya Bay are the following:
\beqa
\tau_d&\sim&2.7\times10^{-6}\ {\rm sec},\no\\
\tau_r&\approx&1\ {\rm hour}\sim3.6\times10^3\ {\rm sec},\no\\
\tau_e&\approx&621\ {\rm days}\sim5.4\times10^7\ {\rm sec}.
\eeqa
The relevant time scales at KamLAND are the following:
\beqa
\tau_d&\sim&6\times10^{-4}\ {\rm sec},\no\\
\tau_r&\approx&10\ {\rm days}\sim8.6\times10^5\ {\rm sec},\no\\
\tau_e&\approx&10\ {\rm years}\sim3\times10^8\ {\rm sec}.
\eeqa
The relevant time scales at JUNO are the following:
\beqa
\tau_d&\sim&1.8\times10^{-4}\ {\rm sec},\no\\
\tau_r&\approx&2.8\ {\rm hours}\sim10^4\ {\rm sec},\no\\
\tau_e&\approx&6\ {\rm years}\sim2\times10^8\ {\rm sec}.
\eeqa
%
\subsection{Solar neutrinos: Super-K and SNO}
Super-K and SNO measure the flux of solar $\nu_e$ neutrinos with event rate of 15/day \cite{Yoo:2003rc} and 10/day \cite{Tolich:2011zza}, respectively. The survival probability $P_{ee}^0$ is given by
\beq\label{eq:peesno}
P_{ee}^0=\sin^2\theta_{12}.
\eeq
The relevant time scales are the following:
\beqa
\tau_r&\approx&1\ {\rm day}\sim8.6\times10^4\ {\rm sec},\no\\
\tau_e&\approx&10\ {\rm years}\sim3\times10^8\ {\rm sec}.
\eeqa
Due to the MSW effect, solar neutrinos propagate as the $\nu_2$ mass state and, consequently, the scale $\tau_d$ is irrelevant here.\\
We summarize the time scales of the relevant experiments in Table \ref{tab:experiments}.
\begin{table}
 \begin{center}
  \caption{The relevant experimental parameters. Time scales are given in units of second. Notice that $\tau_r$ is a rough approximation since it varies for different energy bins. $N_\nu$/day is the unoscillated event rate. }
   \begin{tabular}{cccccc}
   \hline
   & $\tau_d$ & $\tau_r$ & $\tau_e$ & $N_\nu$/day & $P_{\alpha\beta}$ \\ \hline
   Daya Bay & $2.7\times10^{-6}$ & $4\times10^3$ & $5\times10^7$  & 800 & $P_{\bar e\bar e}$ \\
   DUNE     & $4.3\times10^{-3}$ & $1\times10^6$ & $2\times10^8$  & 16 & $P_{\mu\mu},P_{\mu e}$ \\
   HK     & $1.0\times10^{-3}$ & $9\times10^4$ & $9\times10^7$  &  56  & $P_{\mu\mu},P_{\mu e}$ \\
   JUNO     & $1.8\times10^{-4}$ & $1\times10^4$ & $2\times10^8$  & 83& $P_{\bar e\bar e}$ \\
   KamLAND  & $6.0\times10^{-4}$ & $9\times10^5$ & $3\times10^8$ & 2 & $P_{\bar e\bar e}$ \\
   SK, SNO  &                    & $9\times10^4$ & $3\times10^8$  & 10  & $P_{ee}$ \\
   
   \hline
  \end{tabular}
\label{tab:experiments}
 \end{center}
\end{table}

\section{Time-averaged modulations ($\tau_\phi<\tau_r$)}
\label{sec:timeaveraged}
If the ULDM field is fast oscillating, namely with a period that is shorter than the time window over which the data is integrated, an averaged effect can still be probed by experiments. The effect of $\eta_\theta\neq0$ is that experiments measure an effective mixing angle, constant in energy, that is different from the true value of the angle. The effect of $\eta_\Delta\neq0$ is energy smearing of the probability $P_{\mu e}(E)$, competing with the effect of finite energy resolution of the experiment\cite{Krnjaic:2017zlz,Dev:2020kgz}. In this section we analyze the current status and existing bounds as well as the possible future sensitivity to fast oscillations for each of the three mixing angles and the two mass-squared differences separately.

\subsection{Mixing angles\label{Mixing angles}}
We assume here that the dominant effect is time variation in one of the three mixing angles [see Eq.~(\ref{eq:thetahat})]:
\beq\label{eq:thetaij}
\hat\theta_{ij}=\theta_{ij}+\eta_{\theta_{ij}}\sin(m_\phi t).
\eeq
Experiments that measure vacuum oscillations extract an averaged value of $\sin^22\hat\theta_{ij}$:
\beqa\label{eq:fasttij}
\langle\sin^22\hat\theta_{ij}\rangle&=&
\frac{1}{\tau_\phi}\int_0^{\tau_\phi}dt\ \sin^2[2\theta_{ij}+2\eta_{\theta_{ij}}\sin(m_\phi t)]\no\\
&=&\frac12\left[1-\cos(4\theta_{ij})J_0(4\eta_{\theta_{ij}})\right],
\eeqa
where $J_0$ is the zeroth Bessel function of the first kind, fulfilling $0\leq J_0(x)\leq1$, with $J_0(x)=1$ only for $x=0$. As is apparent  from Eq.(\ref{eq:fasttij}), for $\eta_{\theta_{ij}}\neq0$, $\langle\sin^22\hat\theta_{ij}\rangle$ can assume neither the minimal value of zero nor the maximal value of one. The larger $\eta_{\theta_{ij}}$, the further $\langle\sin^22\hat\theta_{ij}\rangle$ is removed from these limiting values. We can then use the extracted values of the three mixing angles $\theta_{ij}$ to put upper bounds on the respective $\eta_{\theta_{ij}}$ values.

Using the range quoted in Ref.~\cite{Zyla:2020zbs} for $\sin^2\theta_{23}$, we obtain
\beq
\langle\sin^22\hat\theta_{23}\rangle=0.992^{+0.006}_{-0.009}.
\eeq
This leads, via Eq. (\ref{eq:fasttij}), to the upper bound
\beq
\eta_{\theta_{23}}<0.09.
\eeq
This bound was obtained previously in Ref.~\cite{Krnjaic:2017zlz}.
Similarly, using the range quoted in Ref.~\cite{Zyla:2020zbs} for $\sin^2\theta_{12}$, we obtain
\beq
\langle\sin^22\hat\theta_{12}\rangle=0.851\pm0.020.
\eeq
This leads, via Eq. (\ref{eq:fasttij}), to the upper bound
\beq
\eta_{\theta_{12}}<0.29.
\eeq
This bound is new.\\
Using the range quoted in Ref.~\cite{Zyla:2020zbs} for $\sin^2\theta_{13}$, we obtain
\beq
\langle\sin^22\hat\theta_{13}\rangle=0.0853\pm0.0027.
\eeq
This leads, via Eq. (\ref{eq:fasttij}), to the upper bound
\beq
\eta_{\theta_{13}}<0.21.
\eeq
This bound was obtained previously in Ref.~\cite{Krnjaic:2017zlz}.\\
Notice that while the method in this section is suited to derive bounds on $\eta_\theta$, it is not designed to give a proof for the existence of such amplitude. In section \ref{Time averaged CPV} we demonstrate how these modulations can be probed directly.

\subsection{Mass-squared differences \label{Mass-squared differences}}
\begin{figure}[t]
 \begin{center}
  \includegraphics[width=0.53\textwidth]{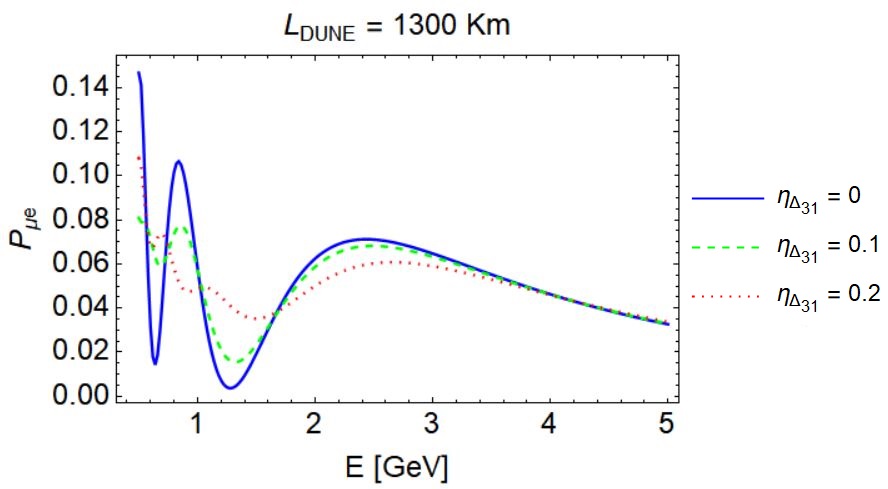} \
 \end{center}
 \caption{Energy smearing effects on $P_{\mu e}(E)$ at DUNE due to $\eta_{\Delta_{31}}$. The effects of the finite energy resolution are not included.}
\label{fig:dune_eta}
\end{figure}

We assume here that the dominant effect is time variation in only one of the two mass-squared differences [see Eq.~(\ref{eq:deltahat})]:
\beq\label{eq:deltaij}
\Delta\hat m^2_{ij}=\Delta m^2_{ij}\left[1+2\eta_{\Delta_{ij}}\sin(m_\phi t)\right].
\eeq
Experiments that measure vacuum oscillations extract an averaged value:
\beq\label{eq:fastdij}
\langle\sin^2[\Delta\hat m^2_{ij}L/(4E)]\rangle&=&
\frac{1}{\tau_\phi}\int_0^{\tau_\phi}dt\ \sin^2\left\{x_{Eij}[1+2\eta_{\Delta_{ij}}\sin(m_\phi t)]\right\}\no\\
&=&\sin^{2}\left(x_{Eij}\right)+2x_{Eij}^{2}\eta_{\Delta_{ij}}^{2}\cos\left(2x_{Eij}\right)+{\cal O}\left(x_{Eij}^{4}\eta_{\Delta_{ij}}^{4}\right).
\eeq
Thus, $\eta_{\Delta_{ij}}$ produces an energy smearing effect which adds to that of the experimental energy resolution. Specifically, an experiment with energy resolution of $X$ is sensitive to $\eta_{\Delta_{ij}}>\frac12 X$. The effect is most significant around local extrema of $P_{\mu e}(E)$: It raises the measured $P_{\mu e}$ at the minima and decreases $P_{\mu e}$ at the maxima. These effects are illustrated for the case of DUNE in Fig.~\ref{fig:dune_eta}.

\subsubsection{$\Delta m^2_{21}$}
\label{sec:Dm21_Kamland}
\begin{figure}[ht]
 \begin{center}
\includegraphics[width=0.46\textwidth]{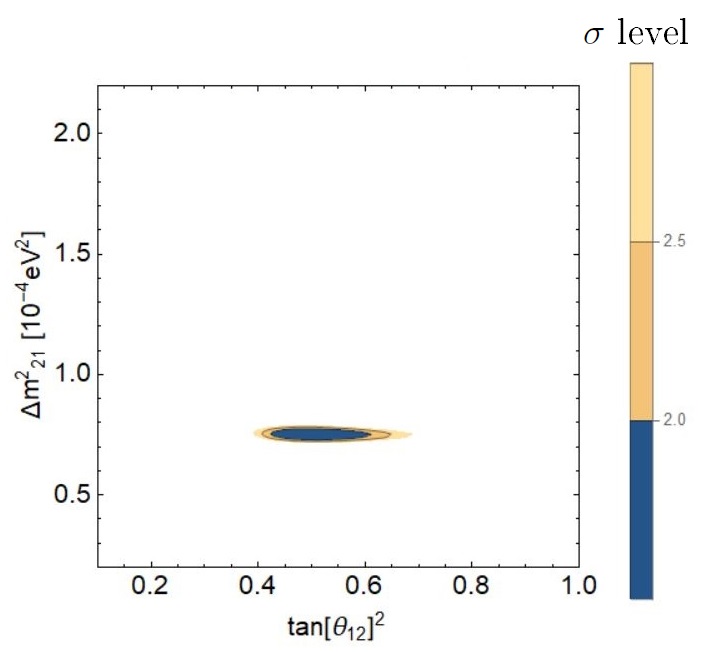}\hfill
\includegraphics[width=0.46\textwidth]{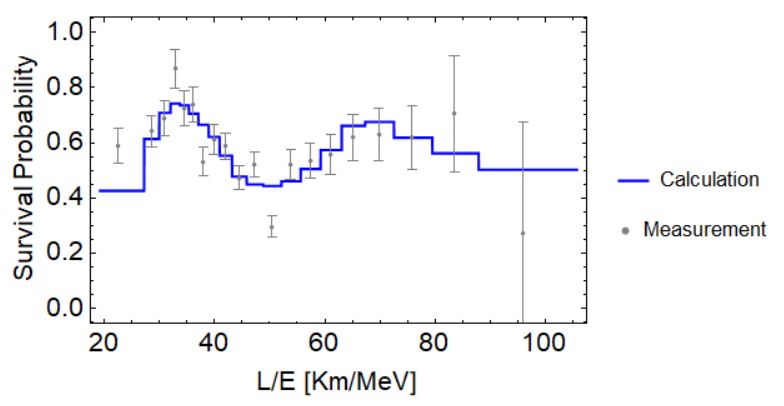}
  \caption{Our fit to the KamLAND measurement reported in Ref.~\cite{Gando:2013nba}:
  \textit{(Left)} Allowed region in the $(\Delta m^2_{21},\tan^2\theta_{12})$ plane.
  \textit{(Right)} The measured $P_{\bar e\bar e}(E)$ and its theoretical value with the best fit parameters.
  }
  \label{fig:kamland_d12t12}
 \end{center}
\end{figure}

\begin{figure}[t]
 \begin{center}
  \includegraphics[width=0.53\textwidth]{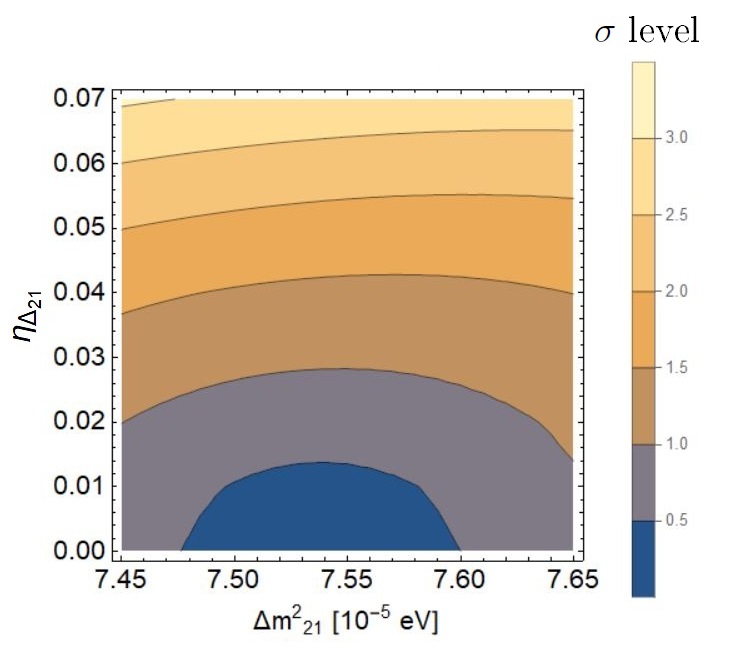} \
 \end{center}
 \caption{Allowed region in the $(\Delta m^2_{21},\eta_{\Delta_{12}})$ plane according to the KamLAND measurement. The other neutrino parameters are fixed at their central values.}
\label{fig:kamlandeta12}
\end{figure}

We now use the extracted value of $P_{\bar e\bar e}(E)$ at the KamLAND experiment to put an upper bound on $\eta_{\Delta_{21}}$.
At KamLAND the energy resolution is $6.4\%$ at $E\sim1$ MeV \cite{Gando:2013nba}, and could thus potentially be sensitive to $\eta_{\Delta_{21}}\sim0.032$ \cite{Krnjaic:2017zlz}. Statistical uncertainties weaken the sensitivity of the experiment, while energy bin correlations may strengthen it. Consequently, to extract the bound from the KamLAND data, we developed  the following procedure:
\begin{enumerate}
\item We calculated the expected spectrum of $e$-events in the case of no oscillations, using the fluxes and distances of the nearest 21 reactors.
\item We calculated the energy-dependent survival probabilities for sets of  values of $(\Delta m^2_{21},\theta_{12})$.
\item We subsequently performed a $\chi^2$ test to find the central values of the parameters and their uncertainties:
\beq
\chi^2=\sum_{i={\rm bins}}\left[\frac{P_{\rm calc}(E_i,\Delta m^2_{21},\theta_{12})-P_{\rm meas}(E_i)}{\sigma(E_i)}\right]^2.
\eeq
The allowed region in the $\Delta m^2_{21}-\tan^2\theta_{12}$ plane is shown in Fig. \ref{fig:kamland_d12t12} (left). We reproduce the central values quoted by KamLAND, but with somewhat smaller uncertainties. The measured $P_{\bar e\bar e}$ and the calculated best fit are shown in Fig. \ref{fig:kamland_d12t12} (Right).
\item We fixed $\theta_{12}$ to the value measured by solar neutrino experiments, and varied the values of $\eta_{\Delta_{21}}$ and $\Delta m^2_{21}$. The resulting allowed region is shown in Fig. \ref{fig:kamlandeta12}.
\end{enumerate}
We finally obtain
\beq\label{eq:kldelta21}
\eta_{\Delta_{21}}<0.05\ ({\rm for}\ 10^{-22}\ {\rm eV}\lsim m_\phi\lsim6\times10^{-12}\ {\rm eV}).
\eeq
This bound is somewhat weaker than the bound quoted in Ref.~\cite{Krnjaic:2017zlz} due to our more detailed treatment of the statistical errors in the KamLAND measurement.

\subsubsection{$\Delta m^2_{31}$}
At present, the case of time-averaged modulations of $\Delta m^2_{31}$ is not directly constrained by experiments. Future experiments, such as JUNO \cite{An:2015jdp} are, however, likely to yield strong constraints  on $\eta_{\Delta_{31}}$, see Ref. \cite{Dev:2020kgz}.

\subsection{CP violation}
\label{sec:cpv}
A CP violating perturbation to neutrinos caused by ULDM, which is modulated in a time averaged manner, can produce a unique signal in neutrino oscillation experiments. 
In general, the matrix $y$ of Eq. (\ref{eq:defzy}) and, therefore, the matrix $\hat y$ of Eq. (\ref{eq:defmyhat}) are complex. Consequently, the CP violation in neutrino oscillations will also exhibit time dependent modulations. It is difficult, however, to construct a model where the only neutrino-related parameter that is affected is the CP violating phase of the leptonic mixing matrix. Yet, to isolate the effects of the time modulations on CP violation, we introduce a toy model that has two interesting features:
\begin{itemize}
\item The matrix $z$ of Eq. (\ref{eq:defzy}) is real, so that the matrix $y$ is the only source of CP violation in the lepton sector.
\item The only non-vanishing entries in $y$ are off-diagonal and purely imaginary, so that the modulation of the neutrino-related CP conserving parameters are quadratic in $\eta$.
\end{itemize}
\subsubsection{Toy model}
In the neutrino mass basis, in the absence of the $\phi$-field, the mass matrix is given by diag$(m_1,m_2,m_3)$. In our toy model, the $\phi$ field introduces small, purely imaginary contributions in the $12$ and $21$ entries:
\begin{equation}
U_{0}^{T}MU_{0}=\left(\begin{array}{ccc}
m_{1} & i\mu & 0\\
i\mu & m_{2} & 0\\
0 & 0 & m_{3}
\end{array}\right).\label{Neta}
\end{equation}
The choice of modulating the 12 and 21 entries is expected to be subject to weaker constraints than the other off-diagonal entries. Here $U_{0}$ is the standard PMNS matrix with $\delta_{\text{CP},0}=0$. The $\mu$ parameter is small in the sense that $|\epsilon|\ll1$, where
\begin{align}
\epsilon\equiv-\frac{\mu}{m_{1}+m_{2}}.
\end{align}
In the presence of the $\mu$-terms, the PMNS matrix $U_0$ needs to be replaced by an $\epsilon$-dependent matrix $U$, which satisfies
\begin{equation}
U^{\dagger}M^{\dagger}MU=D,
\end{equation}
where $D$ is the diagonal mass-squared matrix. We obtain
\begin{equation}
U =U_{0}V,\ \ \
V=\left(\begin{matrix}1-\frac{1}{2}\epsilon^{2} & i\epsilon & 0\\
i\epsilon & 1-\frac{1}{2}\epsilon^{2} & 0\\
0 & 0 & 1
\end{matrix}\right)+{\cal O}\left(\epsilon^{3}\right).
\end{equation}
The corresponding CP violating Jarlskog invariant is proportional to $\epsilon$:
\beq
J  =\frac{\epsilon}{4}\cos\hat{\theta}_{13}\sin2\hat{\theta}_{23}\sin2\hat{\theta}_{13},
\eeq
which gives, in the standard parametrization,
\begin{equation}
\delta_{\text{CP}}=\frac{\epsilon}{\cos\theta_{12}\sin\theta_{12}}+{\cal O}\left(\epsilon^{3}\right).
\end{equation}
Recall that we use the notation $\hat{a}_{ij}$ for a mass or mixing parameter with the presence of the perturbation, and $a_{ij}$ without it. Redefining the angles
\begin{align}
\hat{\theta}_{12} & =\theta_{12}+\frac{\epsilon^{2}}{\tan2\theta_{12}},\\
\hat{\theta}_{13} & =\theta_{13},\\
\hat{\theta}_{23} & =\theta_{23},
\end{align}
brings $U$ to its standard parametrization. Also, the mass-squared differences are shifted via
\begin{align}
\Delta \hat{m}_{21}^{2} & =\Delta m_{21}^{2}\left(1+2\epsilon^{2}\right)\nonumber\\
\Delta \hat{m}_{31}^{2} & =\Delta m_{31}^{2}-2m_{1}\left(m_{1}+m_{2}\right)\epsilon^{2}.\label{eps2}
\end{align}
As expected, the CP violating parameter $\delta_{\text{CP}}$ is modulated at order $\epsilon$, while CP conserving neutrino-related parameters are modulated with $\epsilon^2$.

We now take into account the time-periodic nature of $\epsilon$ by writing:
\begin{equation}
\epsilon=\epsilon_{0}\sin\left(m_{\phi}t\right),
\end{equation}
so we obtain
\beq
\delta_{CP}=\eta_\delta\sin(m_\phi t)+{\cal O}(\eta_{\delta}^3),\ \ \ \eta_\delta=\frac{2\epsilon_0}{\sin2\theta_{12}}.
\eeq
If the modulation period time $\tau_\phi$ is much smaller than the integration time $\Delta t$, the perturbation distorts the neutrino transition and survival probabilities. For example, the CP violating part of $P_{\mu e}$ is averaged out, while the CP conserving part is distorted at order $\epsilon_{0}^2$.\\
Measuring $P_{\mu e}=P_{\bar\mu\bar e}$ would usually be interpreted as a consequence of $\delta_{CP}=0$. Within our model, however, it is a consequence of time-averaging the $\sin(m_\phi t)$ term. We show in section \ref{Time averaged CPV} how the observables of this model cannot be interpreted as a purely CP conserving theory.

\subsubsection{Current bounds on $\epsilon_{0}$}
\begin{figure}[t]
 \begin{center}
  \includegraphics[width=0.53\textwidth]{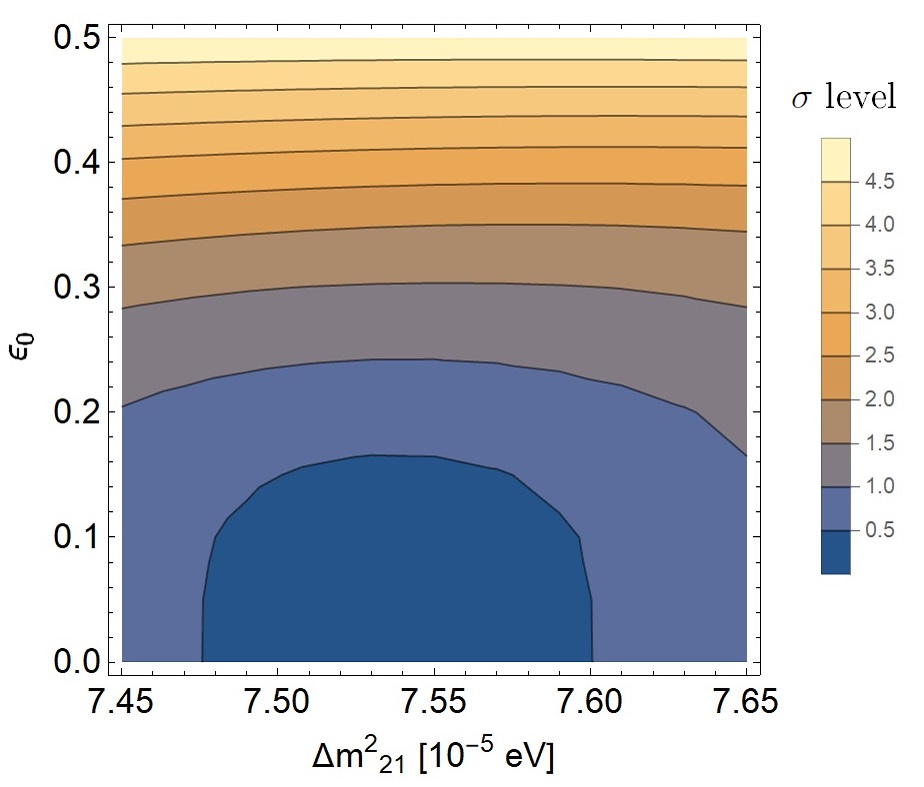} \
 \end{center}
 \caption{Allowed region in the $\Delta m^2_{21}-\epsilon_0$ plane extracted from the KamLAND measurement. The other neutrino parameters are kept fixed at their central values.}
\label{fig:KamLAND_e12}
\end{figure}

Bounds arising from fast $\theta_{12}$ modulation can also bound $\epsilon_{0}$. The strongest current bound comes from KamLAND. We use the method presented in Section \ref{sec:Dm21_Kamland}. Since we are interested in ${\cal O}(\epsilon_{0}^2)$ effects, it can be seen that varying $\epsilon_{0}$ shifts the neutrino parameters that are proportional to $\epsilon_{0}^2$. We need to correct for this shift by changing the vacuum value of these parameters, such that their observed values remain unchanged. We do this by redefining the unperturbed values
\begin{align}
\theta{}_{12} & \rightarrow\theta{}_{12}-\epsilon_0^2/[2\tan\left(2\theta_{12}\right)]\\
\Delta m{}_{21}^{2} & \rightarrow\Delta m{}_{21}^{2}/(2+2\epsilon_{0}^{2}).
\end{align}
We do not shift $\Delta m^2_{31}$ because its modulation is proportional to the lightest neutrino mass $m_1$, and we assume for simplicity that $m_1=0$. The smearing effect due to $\Delta m^2_{21}$ modulation produces a bound from KamLAND. The resulting constraints are shown in Fig. \ref{fig:KamLAND_e12}, and lead to the $2\sigma$ upper bound
\begin{equation}
\epsilon_{0}<0.34\ \Longrightarrow\ \eta_\delta<0.74\,.
\end{equation}
This bound may, however, be over-constraining, because the uncertainties on the neutrino parameters that we derived were approximately $30\%$ smaller than those derived by the KamLAND collaboration. This effect is suppressed in DUNE and HK due to their reduced sensitivity to measure $\Delta m^2_{21}$

\subsubsection{Probing time-averaged CPV\label{Time averaged CPV}}
Despite the cancellation of the CP violating term in the transition probability, a time averaged CP violating perturbation may have a unique imprint in neutrino experiments. We consider a similar perturbation to the one in Eq. \ref{Neta}, but with a complex (and not purely imaginary) value:
\begin{equation}
U_{0}^{T}MU_{0}=\left(\begin{array}{ccc}
m_{1} & \mu e^{i\varphi} & 0\\
\mu e^{i\varphi} & m_{2} & 0\\
0 & 0 & m_{3}
\end{array}\right).
\end{equation}
The correction matrix to the real PMNS matrix is now
\begin{equation}
V=\left(\begin{matrix}1-\frac{1}{2}\epsilon^{2} & \epsilon e^{i\varphi} & 0\\
-\epsilon e^{-i\varphi}  & 1-\frac{1}{2}\epsilon^{2} & 0\\
0 & 0 & 1
\end{matrix}\right)+{\cal O}\left(\epsilon^{3}\right).
\end{equation}
As was demonstrated in Section \ref{Mass-squared differences}, observing peculiar frequencies in the $\frac{L}{E}$ dependence can be explained with real perturbation on the diagonal entries of $M$. We therefore look for peculiar relations between the coefficients of the standard oscillation frequencies. Ignoring additional frequencies that emerge from mass modulations, the survival and transition probabilities can be written as
\begin{align}
P_{\alpha\alpha} & =1-4C_{\alpha\alpha}^{21}\sin^{2}x_{21}-4C_{\alpha\alpha}^{31}\sin^{2}x_{31}-4C_{\alpha\alpha}^{32}\sin^{2}x_{32},\\
P_{\mu e} & =C_{\mu e}^{21}\sin^{2}x_{21}+C_{\mu e}^{31}\sin^{2}x_{31}+C_{\mu e}^{32}\sin x_{31}\sin x_{21}\cos x_{32},
\end{align}
where $x_{ij}=\frac{\Delta m^2_{ij}L}{4E}$, $\alpha = e\text{ or }\mu$, and the CP violating term in $P_{\mu e}$ was averaged out. The coefficients $C^{ij}_{\alpha\beta}$ can be measured directly in neutrino experiments. While it is experimentally difficult in practice to distinguish between $\Delta m_{31}^2$ and $\Delta m_{32}^2$, it can be done in principle, and we assume for this discussion that this is the case. In order to calculate the $C$ coefficients in the model, we need to consider the relevant combination of the lepton mixing matrix elements, and average over $\tau_\phi$ (which leads to the vanishing of any term that is proportional to an odd power of $\epsilon$). To ${\cal O}(\epsilon_0^2)$, we obtain:
\begin{align}
C_{ee}^{21} & =U_{e1}^{2}U_{e2}^{2}+\frac{1}{2}\epsilon_{0}^{2}\left(U_{e1}^{4}-4U_{e1}^{2}U_{e2}^{2}+U_{e2}^{4}\right)
-U_{e1}^{2}U_{e2}^{2}\epsilon_{0}^{2}\cos2\varphi,\nn\\
C_{ee}^{31} & =U_{e3}^{2}U_{e1}^{2}+\frac{1}{2}\epsilon_{0}^{2}U_{e3}^{2}\left(U_{e2}^{2}-U_{e1}^{2}\right),\nn\\
C_{ee}^{32} & =U_{e3}^{2}U_{e2}^{2}-\frac{1}{2}\epsilon_{0}^{2}U_{e3}^{2}\left(U_{e2}^{2}-U_{e1}^{2}\right),\nn\\
C_{\mu\mu}^{21}&=U_{\mu1}^{2}U_{\mu2}^{2}+\frac{1}{2}\epsilon_{0}^{2}
\left(U_{\mu1}^{4}-4U_{\mu1}^{2}U_{\mu2}^{2}+U_{\mu2}^{4}\right)
-U_{\mu1}^{2}U_{\mu2}^{2}\epsilon_{0}^{2}\cos2\varphi,\nn\\
C_{\mu\mu}^{31}&=U_{\mu3}^{2}U_{\mu1}^{2}+\frac{1}{2}\epsilon_{0}^{2}U_{\mu3}^{2}\left(U_{\mu2}^{2}-U_{\mu1}^{2}\right),\nn\\
C_{\mu\mu}^{32}&=U_{\mu3}^{2}U_{\mu2}^{2}+\frac{1}{2}\epsilon_{0}^{2}U_{\mu3}^{2}\left(U_{\mu1}^{2}-U_{\mu2}^{2}\right),\nn\\
C_{\mu e}^{21} & =U_{e2}^{2}U_{\mu2}^{2}+\frac{1}{2}\epsilon_{0}^{2}
\left[U_{e2}^{2}U_{\mu1}^{2}+U_{e1}^{2}U_{\mu2}^{2}-2U_{e2}^{2}U_{\mu2}^{2}+2U_{e1}U_{e2}U_{\mu1}U_{\mu2}
\left(1+\cos2\varphi\right)\right],\nn\\
C_{\mu e}^{31} & =U_{e3}^{2}U_{\mu3}^{2},\nn\\
C_{\mu e}^{32} & =U_{e2}U_{e3}U_{\mu2}U_{\mu3}-\frac{1}{2}\epsilon_{0}^{2}U_{e2}U_{e3}U_{\mu2}U_{\mu3}
+\frac{1}{2}\epsilon_{0}^{2}U_{e1}U_{e3}U_{\mu1}U_{\mu3}\cos2\varphi.
\end{align}
The subscript $0$ was omitted from $U_0$ for simplicity, but here the $U_{ij}$'s refer to the unperturbed real matrix elements. 
The third column of $U$, which is unaffected by the perturbation, can be found in terms of the observed coefficients:
\begin{align}\label{eq:ualpha3}
U_{\mu3}^{2} & =\frac{1+\sqrt{1-4\left(C_{\mu\mu}^{31}+C_{\mu\mu}^{32}\right)}}{2}\\
U_{e3}^{2} & =\frac{C_{\mu e}^{31}}{U_{\mu3}^{2}}\\
U_{\tau3}^{2} & =1-U_{e3}^{2}-U_{\mu3}^{2}.
\end{align}
This is equivalent to finding the unperturbed $\theta_{13}$ and $\theta_{23}$ in the standard parametrization. The third mixing angle $\theta_{12,0}$, the amplitude $\epsilon_{0}$, and the phase of the perturbation $\varphi$ could not be separately extracted from the set of equations, which indicates some degeneracy between them. It is worth mentioning that the solar neutrinos detection probability $P_{ee}$ does not break the degeneracy because it can be calculated in terms of the $C$ coefficients:
\begin{align}
P_{ee} & =U_{e2}^{2}+\frac{1}{2}\epsilon_{0}^{2}\left(U_{e1}^{2}-U_{e2}^{2}\right)=\frac{C_{ee}^{32}}{U_{e3}^{2}}.
\end{align}
The assumption that the ULDM field is constant during the neutrino's propagation is not required for solar neutrinos, because for adiabatic propagation they remain as a mass state $\nu_2$.
Even though the parameters cannot be extracted exactly, we can learn about the nature of the perturbation by defining
\begin{equation}\label{xi_alphabeta}
\xi_{\mu e} =\frac{\left(C_{\mu e}^{32}\right)^2}{C_{\mu e}^{21}C_{\mu e}^{31}}=1-\frac{\left(U_{e2}U_{\mu1}+U_{e1}U_{\mu2}\right)^{2}}{2U_{e2}^{2}U_{\mu2}^{2}}\epsilon_{0}^{2}+{\cal O}\left(\epsilon_{0}^{4}\right).
\end{equation}
Establishing an inequality,  $\xi_{\mu e}<1$, would indicate time-averaged new physics, which modulates elements of the mixing matrix. The inequality by itself can, however, be interpreted as modulation of the mixing angles (Section \ref{Mixing angles}), and not necessarily of the phase $\delta_{\text{CP}}$. 

The coefficient $\frac{\left(U_{e2}U_{\mu1}+U_{e1}U_{\mu2}\right)^{2}}{2U_{e2}^{2}U_{\mu2}^{2}}>0$ can be extracted from measurements to zeroth order in $\epsilon_0$. This approximation is, however, all that we need for our purposes since the coefficient multiplies $\epsilon_0^2$. For example, it can be found using
\begin{align}
U_{e2}^{2}U_{\mu2}^{2} & =C_{\mu e}^{21}+{\cal O}\left(\epsilon_{0}^{2}\right),\\
U_{e2} & =U_{e3}\sqrt{\frac{C_{ee}^{21}}{C_{ee}^{31}}}+{\cal O}\left(\epsilon_{0}^{2}\right),\\
U_{\mu1} & =U_{\mu3}\sqrt{\frac{C_{\mu\mu}^{21}}{C_{\mu\mu}^{32}}}+{\cal O}\left(\epsilon_{0}^{2}\right),
\end{align}
and unitarity. Then $\epsilon_0$, the amplitude of the perturbation,  can be obtained from the deviation of $\xi_{\mu e}$ from $1$. In order to check whether $\delta_{\text{CP}}$ is being modulated, we also define:
\begin{align}
\label{xi_alphaalpha}\xi_{\alpha\alpha} & =\sqrt{\frac{C_{\alpha\alpha}^{21}C_{\alpha\alpha}^{31}}{C_{\alpha\alpha}^{32}}}
+\sqrt{\frac{C_{\alpha\alpha}^{21}C_{\alpha\alpha}^{32}}{C_{\alpha\alpha}^{31}}}
+\sqrt{\frac{C_{\alpha\alpha}^{31}C_{\alpha\alpha}^{32}}{C_{\alpha\alpha}^{21}}}
=1-\left(1-2U_{\alpha3}^{2}\right)\epsilon_{0}^{2}\cos^{2}\varphi,
\end{align}
where $\alpha$ can be $e$ or $\mu$ for reactor or accelerator experiments. Unlike $\xi_{\mu e}$ which is sensitive to $\epsilon_0$, the absolute value of the perturbation, $\xi_{\alpha\alpha}$ is sensitive to $\epsilon_0\cos\varphi$, its real part. The coefficient $\left(1-2U_{\alpha3}^{2}\right)$ can be calculated here in terms of the physical coefficients, see Eq. (\ref{eq:ualpha3}). Thus, combining the measurements of $1-\xi_{\mu e}$ and $1-\xi_{\alpha\alpha}$ allows one to extract the phase of the perturbation $\varphi$. A non vanishing phase means that $\delta_{\text{CP}}$ is being modulated in a time-averaged manner.

Using error propagation on the current measured values and uncertainties of the mixing angles, we estimate the uncertainty of $\xi$ to be of order of $\sim0.1$, which is the same order as the prefactor of $\epsilon_{0}^2$ in Eq.~(\ref{xi_alphabeta}). A more careful calculation is expected to yield even worse sensitivity to $\epsilon_{0}$ due to the difficulty to separate $\Delta m_{31}^2$ from $\Delta m_{32}^2$. Current neutrino experiments are, thus, not sensitive enough to probe time averaged CPV, but in principle it may be possible in the future. This would require experimental capability to have energy resolution better than $\frac{\Delta m_{21}^{2}}{\Delta m_{32}^{2}}\approx3\%$, which will allow the separation of $\Delta m_{32}^{2}$ from $\Delta m_{31}^{2}\,$.


\section{Time-resolved modulations ($\tau_\phi>\tau_r$)}
\label{sec:timeresolved}
If the ULDM field is slowly oscillating, namely with a period that is longer than the time window over which the data is integrated, the time dependence of $P_{\alpha\beta}(E)$ could, in principle, be probed by experiments. Given values of $m_\phi$ and $\eta$ (which set the value of $\epsilon_{\alpha\beta}(E)$ in Eq.~(\ref{P0+epsilon})), we can calculate the confidence level (CL) at which a given experiment is expected to observe a signal (see Appendix \ref{app:slow} for details):
\beq\label{eq:clme}
{\rm CL}=\left[1-\exp\left(-\frac14 \frac{N_\nu^{\rm tot}(E){\rm sinc}^2(m_\phi\Delta t/2)
|\langle\epsilon_{\alpha\beta}(E)\rangle|^2}
{\langle P^0_{\alpha\beta}(E)\rangle(1-\langle P^0_{\alpha\beta}(E)\rangle)}\right)\right]^{\frac{\tau_e}{\Delta t}}.
\eeq
Sensitivity to time modulation in neutrino experiments requires a very high rate of neutrino events.
The relevant parameters of the experiments that can probe slow oscillations are given in Table \ref{tab:experiments}.

\subsection{Current bounds}
Due to the MSW effect, the solar neutrinos reach Earth in the mass eigenstate $\nu_2$. Thus, experiments measuring the solar $\nu_e$ flux are sensitive to $|U_{e2}|^2=\sin^2\theta_{12}\cos^2\theta_{13}$. Since the solar neutrinos make their way from Sun to Earth as mass eigenstates, there are no neutrino oscillations, and one should not use Eq.~(\ref{eq:clme}) to deduce the experimental sensitivity. Instead, Eq. (\ref{eq:thetahat}) implies
\begin{align}
\sin^{2}\hat{\theta}_{12}\left(t\right) & =\sin^{2}\theta_{12}+\eta_{\theta_{12}}\sin2\theta_{12}\sin\left(m_{\phi}t\right)+{\cal O}\left(\eta_{\theta_{12}}^{2}\right),\nn\\
\cos^{2}\hat{\theta}_{13}\left(t\right) & =\cos^{2}\theta_{13}-\eta_{\theta_{13}}\sin2\theta_{13}\sin\left(m_{\phi}t\right)+{\cal O}\left(\eta_{\theta_{13}}^{2}\right).
\end{align}
An experiment that is sensitive to modulations larger than a fraction $X$ of $|U_{e2}|^2$ and observes no such effect, will therefore put the bounds
\begin{align}
\eta_{\theta_{12}} & \lesssim\frac{1}{2}\tan\theta_{12}X\backsimeq0.33X,\nn\\
\eta_{\theta_{13}} & \lesssim\frac{1}{2\tan\theta_{13}}X\backsimeq3.35X,
\end{align}
where we use Ref. \cite{Zyla:2020zbs} for the values of the mixing angles.

The Super-K and SNO experiments searched for periodic time variations in their signal (beyond the effect of the eccentricity of the Earth's orbit), and found none over 10\%. Thus,
\begin{align}
\eta_{\theta_{12}} & <0.03,\\
\eta_{\theta_{13}} & <0.3.
\end{align}
The bound on $\eta_{\theta_{13}}$ is weaker than the one from time-averaged modulation. The bound on $\eta_{\theta_{12}}$ is consistent with the result given in Ref.~\cite{Berlin:2016woy}. The bound holds for
\beq
10\ {\rm minutes}<\tau_\phi<10\ {\rm years}\ \Longrightarrow\ 10^{-23}\ {\rm eV}<m_\phi<7\times10^{-18}\ {\rm eV}.
\eeq
The limit on the largest time scale corresponds to the value of $\tau_e$ of these experiments. Even though $\tau_r=1$ day, a better resolution for shorter period times between $1$ day and $10$ minutes was obtained using the unbinned Rayleigh method \cite{Collaboration:2009qz}. Eq. (\ref{eq:clme}) however, does not consider this method, and therefore loses its sensitivity when $\tau_\phi<\tau_r$.

To analyze periodic time modulation due to $\theta_{13}$, $\Delta m^2_{31}$ and $\Delta m^2_{21}$, one can search for time variations in $P_{\bar e\bar e}$ measured in short-baseline reactor neutrino experiments. The corresponding expression for $P_{\bar e\bar e}$ is given in Eq. (\ref{eq:peedayabay}).

The Daya Bay experiment collected 800 unoscillated daily events in the near detector \cite{An:2016ses} and searched for time variations in $P_{\overline e\overline e}$ \cite{Adey:2018qsd}. We use $N_\nu^{\rm tot}\approx10^5$ (in the bin $E\in[0.7,2]$ MeV), $t_e=621$ days and $t_r=1$ hour. We infer the expected CL as a function of $m_\phi$ and each of $\eta_{\theta_{13}}$, $\eta_{\Delta_{31}}$ and $\eta_{\Delta_{21}}$ using Eq. (\ref{eq:clme}). Given that no signal of time variation was observed \cite{Adey:2018qsd}, we obtain, for $10^{-22}\text{ eV} \lesssim m_{\phi}\lesssim7\times10^{-18}\text{ eV}$,
\begin{align}
\eta_{\theta_{13}} & <0.013,\\
\eta_{\Delta_{31}} & <0.08,\\
\eta_{\Delta_{21}} & <0.3.
\end{align}

\begin{figure}[t]
 \begin{center}
  \includegraphics[width=0.53\textwidth]{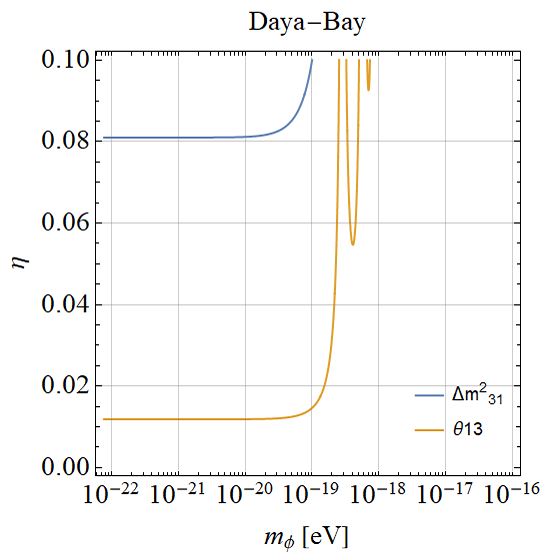} \
 \end{center}
 \caption{$1\sigma$ contours for the sensitivity of Daya-Bay to detect ULDM signal due to modulation of $\Delta m^2_{31}$ (blue), and $\theta_{13}$ (orange).}
\label{fig:DBsensitivity}
\end{figure}

The sensitivity of Daya Bay to time modulation of $\theta_{13}$ and $\Delta m^2_{31}$ is shown in Fig. \ref{fig:DBsensitivity}. The bound on $\eta_{\Delta_{21}}$ is weaker than the bound given in Eq. (\ref{eq:kldelta21}) from KamLAND on time-averaged modulations.\\
We did not extract bounds from KamLAND for time-resolved modulations. First, the analysis is more complicated due to the different distances between the detector and the reactors. Second, the statistics is relatively low. Thus, we expect the KamLAND bounds to be weaker.

\subsection{Future sensitivities}
\begin{figure}[t]
 \begin{center}
  \includegraphics[width=1\textwidth]{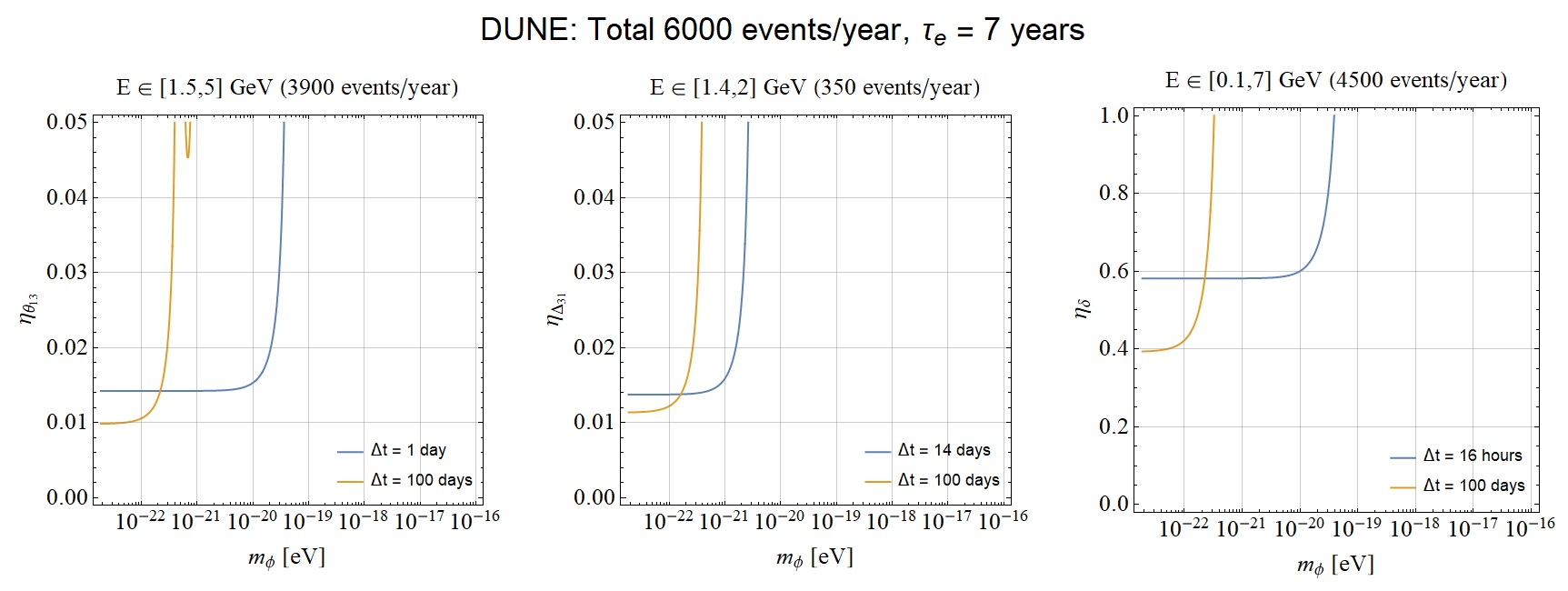} \
 \end{center}
 \caption{$1\sigma$ contours for the expected sensitivity of DUNE to detect ULDM signal due to modulation of $\theta_{13}$ (left), $\Delta m^2_{31}$ (middle), and $\delta_{CP}$ (right) via $P_{\mu e}$ measurements. The shorter integration time (blue graph) in each figure was set such that it would include $10$ unoscillated events, and the longer integration time (orange graph) is fixed at $100$ days.}
\label{fig:DUNEsensitivity}
\end{figure}

The future DUNE experiment will measure $P_{\mu\mu}$ and $P_{\mu e}$.  Fig. \ref{fig:DUNEsensitivity} shows the expected sensitivities of DUNE to modulations of the mixing angle $\theta_{13}$ (left), the mass-squared difference $\Delta m^2_{31}$ (middle), and the CP-violating phase $\delta_{CP}$ (right). We consider the total run time for the experiment to be 7 years with an event rate of 6000 unoscillated events/year. We have checked the sensitivity to these parameters in both $P_{\mu\mu}$ and $P_{\mu e}$, and the latter yields better sensitivity for all of them. For each of the three parameters, a specific energy bin has been selected. We show the $1\sigma$ contours for two possible choices of integration time. The first choice fixes the integration time to contain approximately 10 unoscillated events, such that the calculation is valid (see Appendix \ref{app:slow}). The second choice fixes the integration time to 100 days. The sensitivity we obtain in the case of DUNE to $\eta_{\Delta_{31}}$ is slightly stronger than the sensitivity quoted in \cite{Dev:2020kgz}. This difference might be due to different energy bins that are being probed, or due to a different method to look for modulations, or due to different sensitivities of survival and transition probabilities.

\begin{figure}[]
	\begin{center}
		\includegraphics[width=1\textwidth]{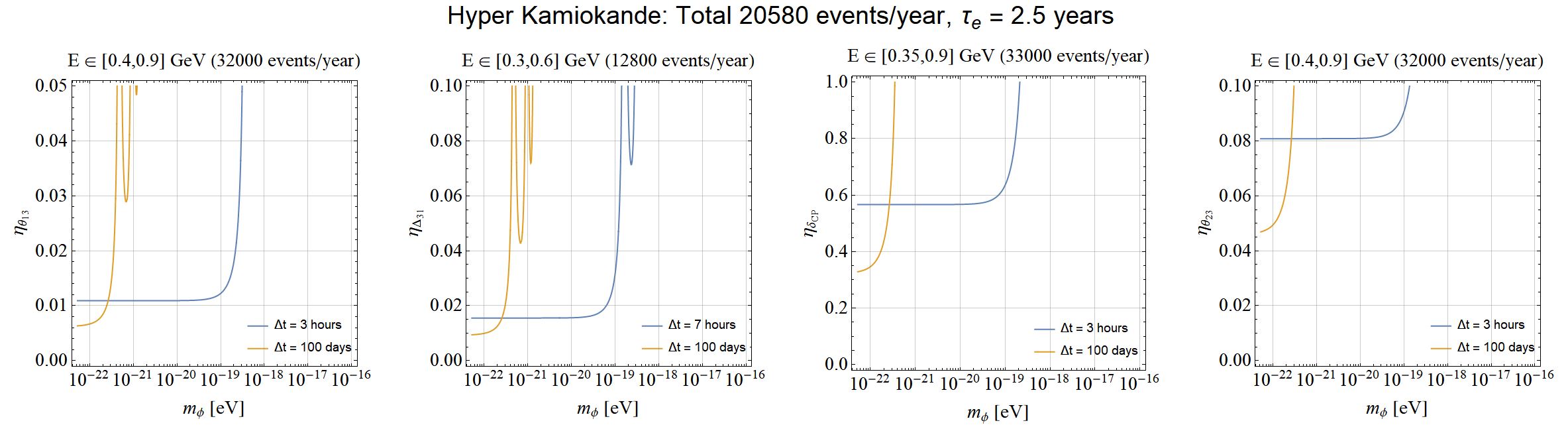} \
	\end{center}
	\caption{$1\sigma$ contours for the expected sensitivity of HK to detect ULDM signal due to modulation of $\theta_{13}$ (left), $\Delta m^2_{31}$ (second from left), $\delta_{CP}$ (second from right), and $\theta_{23}$ (right) via $P_{\mu e}$ measurements. The shorter integration time (blue graph) in each figure was set such that it would include $10$ unoscillated events, and the longer integration time (orange graph) is fixed at $100$ days.}
	\label{fig:T2HKsensitivity}
\end{figure}

The future Hyper Kamiokande, HK, experiment is similar to DUNE in the sense that they have similar $L/E_\nu$, and they both measure $P_{\mu\mu}$ and $P_{\mu e}$. However, the statistics of HK are superior to those of DUNE, and therefore it produces better sensitivities to modulation amplitudes for a shorter running time. The expected sensitivities in HK are presented in Fig. \ref{fig:T2HKsensitivity}.

\begin{figure}[]
 \begin{center}
  \includegraphics[width=1\textwidth]{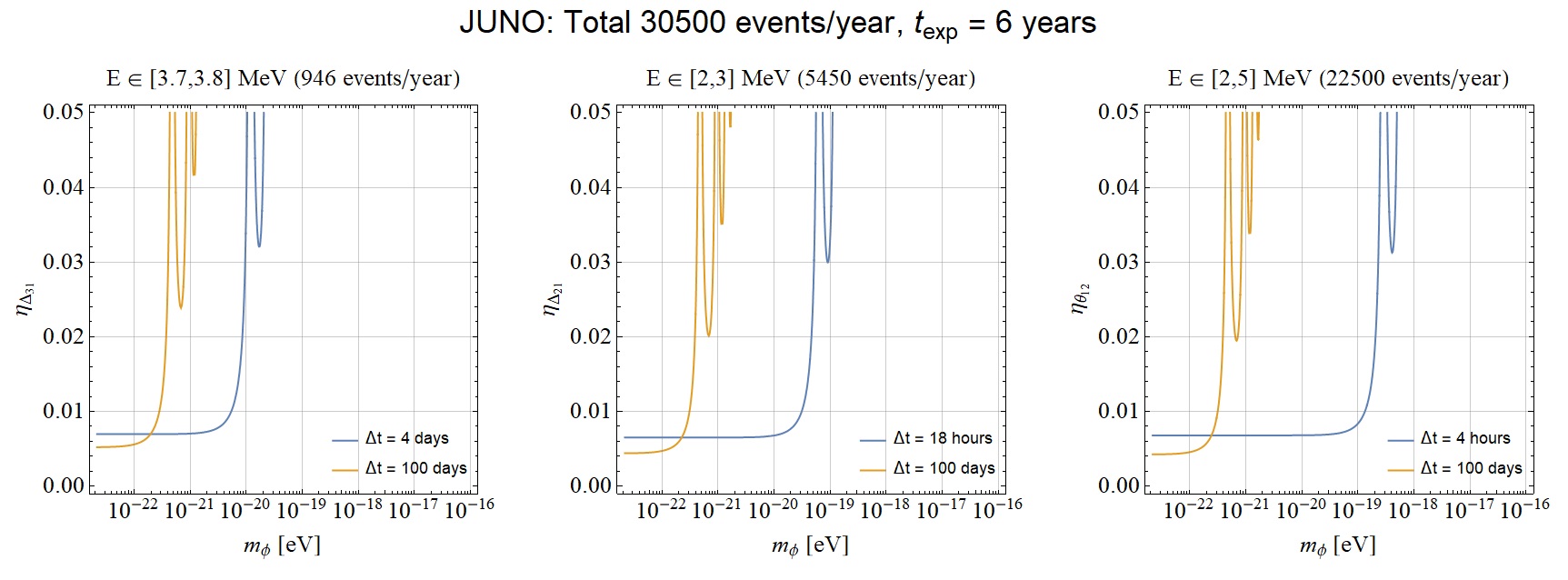} \
 \end{center}
 \caption{$1\sigma$ contours for the expected sensitivity of JUNO to detect ULDM signal due to modulation of  $\Delta m^2_{21}$ (left), $\Delta m^2_{31}$ (middle), and $\theta_{12}$ (right). The shorter integration time (blue graph) in each figure was set such that it would include $10$ unoscillated events, and the longer integration time (orange graph) is fixed at $100$ days.}
\label{fig:JUNOsensitivity}
\end{figure}

The future JUNO experiment will measure $P_{\bar e\bar e}$, and therefore is not sensitive to the $\delta_{\text{CP}}$ parameter. The total run time for the experiment is taken to be 6 years, with unoscillated 30500 events/year \cite{Giaz:2018gdd}. Fig. \ref{fig:JUNOsensitivity} shows the expected sensitivities of JUNO for modulation of $\theta_{13}$, $\Delta m^2_{31}$, and $\theta_{12}$.


\section{Other phenomenological implications of ULDM-neutrino coupling}
\label{sec:pheno}
At the quantum level, the Lagrangian of Eq.~\eqref{eq:defzy} induces a potential for the Higgs and the ULDM fields.
First setting the Higgs field to its VEV, we find the following dominant, tadpole, one-loop contributions to the Coleman-Weinberg potential:
$V^{\rm CW}_\phi\sim \hat y \Lambda_\nu^2  m_\nu \,\phi/(16 \pi^2)+{\rm h.c.} \,,$ with $\Lambda_\nu$ being the cutoff that characterizes the NP scale that regularizes the loop divergence. Adding the bare mass term, the effective potential for $\phi$ is
\begin{equation}
V_\phi=m_\phi^2 \phi^2+\hat y \Lambda_\nu^2  m_\nu  \,\phi/16 \pi^2+{\rm h.c.}
\end{equation}
It implies that $\phi$ develops a VEV of the order of $\phi_0\sim \hat y \Lambda_\nu^2  m_\nu /(16 \pi^2 m_\phi^2)\,.$ On one hand, we would like to have $\hat y \phi_0/m_\nu\lsim1$, so that its contributions to the neutrino masses are subdominant. On the other hand, to have a visible signal for the neutrino experiments discussed above, we require that $\hat y \phi\gsim 0.1 m_\nu$, which implies $\hat y \gsim 0.1 m_\phi m_\nu/\sqrt{\rho_{\rm DM}}$. Thus, we find:
\begin{equation}
\Lambda_\nu \lesssim \frac{4 \pi  m_\phi}{\hat y}  \lesssim 10^2 \sqrt{\rho_{\rm DM}\over m_\nu^2}
\sim 10{\rm \,eV}\,.
\end{equation}
We learn that, for our scenario to be both natural and experimentally observable, new degrees of freedom that couple to $\phi$ (say sterile neutrinos or sneutrinos) with mass scale of 10\,eV or less are required to be present.

The above tadpole contribution vanishes in cases, considered above, where in flavor space the bare mass matrix is orthogonal to the scalar Yukawa, namely tr\,$\left(m_\nu \hat y\right)=0\,.$
We can, however, obtain a similar naturalness-based bound by simply demanding that the quadratically divergent contribution to the scalar mass is smaller than its physical value: $\delta m_\phi^2\sim \hat y^2 \Lambda_\nu^2 /(16 \pi^2) \lesssim m_\phi^2$. This implies $\Lambda_\nu \lesssim 4 \pi  m_\phi/\hat y\,,$ as in the above.

The above quantum corrections lead to $\phi-H$ mixing. If we reconsider  the contributions to the Coleman-Weinberg potential without setting all the Higgs-field-insertions to their VEV, we find that a trilinear scalar interaction term is induced:
\begin{equation}
V(H^\dagger H \phi)\sim \frac{\hat y \Lambda_\nu^2  m_\nu}{16 \pi^2 v^2} H^\dagger H \phi\,.
\end{equation}
It leads to $\phi-H$ mixing, with the mixing angle given by
\begin{equation}
\theta_{H\phi}\sim {\hat y \Lambda_\nu^2  m_\nu \over  16 \pi^2 v  m_H^2}\gtrsim
{0.1 m_\phi m_\nu\over \sqrt{\rho_{\rm DM}}} \, {\Lambda_\nu^2  m_\nu \over  16 \pi^2 v  m_H^2}\sim 10^{-54}\, {m_\phi\over 10^{-18}\rm\, eV}\times \left( {\Lambda_\nu\over{\rm eV}}\right)^2
 \,.
 \end{equation}
This small mixing is consistent with bounds from the equivalence principle and atomic clocks (see Ref.~\cite{Banerjee:2020kww} for  a recent analysis).

Finally, we comment on CMB bounds on the parameters of our model. In the cosmological context, the ULDM amplitude of oscillation is proportional to $(1+z)^{3/2}$, which leads to a larger effective mass for the neutrinos at earlier time in the evolution of the Universe. Thus, the upper bound on the sum of neutrino masses from the CMB data can constrain the ULDM-neutrino Yukawa interactions \cite{Berlin:2016woy,Krnjaic:2017zlz,Brdar:2017kbt,Dev:2020kgz}.
The CMB bound, $\Sigma m_\nu\lsim 0.2$ eV~\cite{Krnjaic:2017zlz,Brdar:2017kbt}, results in $\eta_{\Delta_{31}}\lesssim 3\times 10^{-2}$, which is stronger than current bounds obtained in neutrino oscillation experiments.
It is important to note, however, that this bound is sensitive to the effect of the DM amplitude on the neutrino masses, but to neither its effects on the mixing angles nor on the CP phase. In the model defined in Eq.~\eqref{Neta}, the DM amplitude modifies the masses only at quadratic order, $\epsilon^2$, see Eq.~\eqref{eps2}, and Eq.~\eqref{eq:hatpmuetheta} for the model independent relations.
We conclude that the current cosmological bound is, in some cases, weaker than the direct ones obtained in this work.

\section{Summary}
\label{sec:summary}
We studied the constraints on ultra-light dark matter (ULDM) from neutrino oscillations. We considered scenarios of time-averaged and time-resolved modulations of the neutrino mass and mixing parameters due to the neutrino couplings to the ULDM field. We constructed a model that demonstrates that there could be a situation where the leading (linear in $\phi$) effect is CP violating, while the modulation of CP conserving parameters is suppressed (at most quadratic in $\phi$). We derived bounds on our model from KamLAND, and showed a unique imprint that a time-averaged CP violation can have on neutrino experiments.

A summary of our results is given in Table \ref{tab:summary}, where the novel bounds derived in our study are given in bold. We note that, in the analysis for time-resolved modulation, we did not look for correlation between energy bins; Doing so may further improve the sensitivities.
\begin{table}
 \begin{center}
  \caption{Current and projected bounds on $\eta$. New bounds are quoted in bold letters. Data are taken from Super-K (SK), SNO, Daya Bay (DB) and KamLAND (KL).}
   \begin{tabular}{l|ccc|ccc|cc|cc|cc}
   \hline
   $\eta$& \multicolumn{3}{c}{$\tau_\phi<\tau_r$} & \multicolumn{3}{c}{$\tau_\phi>\tau_r$} & \multicolumn{2}{c}{JUNO}& \multicolumn{2}{c}{DUNE}& \multicolumn{2}{c}{HK} \\
   & Bound & $P_{\alpha\beta}$ & Exp. & Bound & $P_{\alpha\beta}$ & Exp. & Bound & $P_{\alpha\beta}$& Bound & $P_{\alpha\beta}$ & Bound & $P_{\alpha\beta}$ \\
   \hline
   $\eta_{\theta_{12}}$ & ${\bf0.29}$ & $P_{ee}$ & $\nu_{\rm sol}$ & $0.03$ & $P_{ee}$ & SK,SNO & ${\bf0.004}$ & $P_{\bar e\bar e}$& & & & \\
   $\eta_{\theta_{13}}$ & $0.21$ & $P_{\bar e\bar e}$ & $\nu_{\rm rea}$ & ${\bf0.01}$ & $P_{\bar e\bar e}$ & DB &  &  &${\bf0.01}$ &$P_{\mu e}$ &${\bf0.006}$ &$P_{\mu e}$ \\
   $\eta_{\theta_{23}}$ & $0.09$ & $P_{\mu\mu}$ & $\nu_{\rm atm}$ & $-$ & & &  & & & & ${\bf0.046}$ &$P_{\mu e}$ \\
    $\eta_{\delta}$ & ${\bf0.74}$ & $P_{\bar e\bar e}$ & KL & $-$ & & &  & &${\bf0.4}$ &$P_{\mu e}$ & ${\bf0.33}$ & $P_{\mu e}$ \\
   $\eta_{\Delta_{31}}$ & $-$ &&& ${\bf0.08}$ & $P_{\bar e\bar e}$ & DB & ${\bf0.005}$ & $P_{\bar e\bar e}$ &${\bf0.01}$ &$P_{\mu e}$ &${\bf0.01}$ &$P_{\mu e}$ \\
   $\eta_{\Delta_{21}}$ & $0.05$ & $P_{\bar e\bar e}$ & KL & $0.3$ & $P_{\bar e\bar e}$ & DB & ${\bf0.004}$ &  $P_{\bar e\bar e}$& & & &  \\
   \hline
  \end{tabular}
\label{tab:summary}
 \end{center}
\end{table}

We highlight the following points with regard to the novel bounds:
\begin{itemize}
\item The bound on $\eta_{\theta_{12}}$ from time-averaged modulations: Previous bounds from time-resolved modulations hold only for $m_\phi\lsim7\times10^{-18}$ eV.
\item The bound on $\eta_{\theta_{13}}$ from time-resolved modulations: It is the strongest bound for $m_\phi\lsim7\times10^{-18}$ eV.
\item The bound on $\eta_{\Delta_{31}}$ from time-resolved modulations: This constitutes the first bound on this parameter.
\item The bound on $\eta_\delta$ from time-averaged modulations from KamLAND: This is the first bound on this parameter.
\item The expected sensitivities to various $\eta$'s  of the different neutrino mass and mixing parameters from JUNO, DUNE and HK.
\end{itemize}

\noindent We argued that naturalness considerations imply that, if the scenario we consider is to give an observable signal in neutrino oscillation experiments, new degrees of freedom that interact with the ULDM are required, with a mass scale lower the $10$ eV.

\appendix
\section{The statistical analysis of slow modulations}
\label{app:slow}
Slow modulations, $\tau_r<\tau_\phi<\tau_e$, allow us to split the running time of an experiment into smaller time segments and measure $P_{\alpha\beta}(E)$ in each of them. For neutrinos propagating in a ULDM field, we expect a small time varying perturbation to the transition probability,
\beq
\label{P0+epsilon}P_{\alpha\beta}(E,t)\approx P^0_{\alpha\beta}(E)+\epsilon_{\alpha\beta}(E)\sin(m_\phi t)+{\cal O}(\epsilon_{\alpha\beta}^2),
\eeq
where $\epsilon_{\alpha\beta}\propto\eta$. For example, in the two neutrino case, Eq. (\ref{eq:hatpmue}) yields
\beq\label{eq:epsilonab}
\epsilon_{\mu e}=2\eta_{\theta}\sin2\theta\sin^2x_E
+2\eta_{\Delta}x_E\sin^22\theta\sin2x_E.
\eeq
$P^0_{\alpha\beta}$ is the probability that is measured when integrating over time $\gg\tau_\phi$. To first order in $\eta$, it coincides with the vacuum transition probability.

Let us consider the case that we measure $P_{\alpha\beta}(E_0,t_n)$ by integrating over time interval $\Delta t$ around time $t_n$ and in an energy bin of width $\Delta E$ around $E_0$:
\beqa
\langle P_{\alpha\beta}(E_0,t_n)\rangle&=&\frac{1}{\Delta t\Delta E}
\int_{E_0-\frac{\Delta E}{2}}^{E_0+\frac{\Delta E}{2}}
\int_{t_n-\frac{\Delta t}{2}}^{t_n+\frac{\Delta t}{2}}W(E)P_{\alpha\beta}(E,t)dEdt\no\\
&=&\langle P^0_{\alpha\beta}(E_0)\rangle+{\rm sinc}\left(m_\phi \Delta t/2\right)
\langle\epsilon_{\alpha\beta}(E_0)\rangle\sin(m_\phi t_n),
\eeqa
where $W(E)$ is a normalized weight function fulfilling $\int dEW(E)=1$. This function is calculated by the product of the neutrino flux spectrum and their cross section in the detector. Its role is important for wide enough energy bins such that the event rate is not uniform in energy.

In order to detect, or constrain, a time periodic modulation, it is beneficial to study the Fourier transform of $\langle P_{\alpha\beta}(E,t)\rangle$:
\beqa\label{eq:tildep}
\widetilde P_{\alpha\beta}(E,\omega)&=&\sum_{n=0}^{n_{\rm tb}-1}
\langle P_{\alpha\beta}(E,t_n)\rangle e^{-i\frac{2\pi n}{n_{\rm tb}}\omega}\no\\
&=&\langle P^0_{\alpha\beta}(E)\rangle\sum_{n=0}^{n_{\rm tb}-1} e^{-i\frac{2\pi n}{n_{\rm tb}}\omega}
+{\rm sinc}\left(m_\phi \Delta t/2\right)
\langle\epsilon_{\alpha\beta}(E)\rangle\sum_{n=0}^{n_{\rm tb}-1} \sin(m_\phi t_n)e^{-i\frac{2\pi n}{n_{\rm tb}}\omega},
\eeqa
where $n_{\rm tb}$ is the number of time bins or, equivalently, the number of $P_{\alpha\beta}$ measurements, so that $\tau_e=n_{\rm tb}\Delta t$.

To gain intuition about the best way to discover the periodic modulation of the neutrino transition probability, and the sensitivity of the experiment, we provisionally consider only values of $\omega$ that are integer multiples of the smallest frequency that the experiment can probe,
\beq
\omega_0\equiv \tau_{\rm exp}^{-1}=(n_{\rm tn}\Delta t)^{-1}.
\eeq
Then, Eq. (\ref{eq:tildep}) is simplified considerably:
\beq\label{eq:tpab}
\left|\widetilde P_{\alpha\beta}(E,\omega)/n_{\rm tb}\right|^2=
\left|\langle P^0_{\alpha\beta}(E)\rangle\right|^2\delta_{0,\omega}
+\left|{\rm sinc}\left(m_\phi \Delta t/2\right)
\langle\epsilon_{\alpha\beta}(E)\rangle\right|^2\left(\delta_{m_\phi,\omega}+\delta_{-m_\phi,\omega}\right)/4.
\eeq

Eq. (\ref{eq:tpab}) leads us to expect in the Fourier picture a primary peak at $\omega=0$, which corresponds to the neutrino vacuum oscillations, and two secondary peaks at $\omega=\pm m_\phi$, which correspond to the ULDM field oscillations. In practice, due to the statistical errors in measuring $\langle P_{\alpha\beta}(E_0,t_n)\rangle$, there is also noise on top of the signal in the Fourier picture. In order to detect the $\phi$-field, the secondary peaks must be significantly higher than this noise.

We now set to find the statistical error in the measurement of $\langle P_{\alpha\beta}(E,t_n)\rangle$. We do so to zeroth order in $\epsilon_{\alpha\beta}$. With this approximation, $P_{\alpha\beta}(E,t_n)=P^0_{\alpha\beta}(E)$ and the number of unoscillated events in a given energy bin and time bin is time-independent, $N_\nu(E,t_n)=N_\nu(E)$. Then
\beq
\Delta P_{\alpha\beta}(E,t_n)=\sqrt{\frac{\langle P^0_{\alpha\beta}(E)\rangle(1-\langle P^0_{\alpha\beta}(E)\rangle)}{N_\nu(E)}},
\eeq
where $N_n$ is the number of events in the $t_n$-bin during the time interval $\Delta t$. The standard deviation in $\widetilde P_{\alpha\beta}(E)$ is given by
\beq
\Delta\widetilde P_{\alpha\beta}(E)=\Delta P_{\alpha\beta}(E)\sqrt{n_{\rm tb}/2}.
\eeq
This relation is valid for $N_\nu(E)\gg1$, which is the reason we always make sure there are more than 10 events in the energy bin at each measurement of $P_{\alpha\beta}$. Note that $\Delta\widetilde P_{\alpha\beta}\propto\sqrt{n_{\rm tb}}$ while $\widetilde P_{\alpha\beta}\propto n_{\rm tb}$.

To obtain the CL we should find the probability that a secondary peak in $|\widetilde P_{\alpha\beta}(E,m_\phi)|^2$, with value $|\widetilde P|^2_{\rm max}$, is not just a statistical fluctuation of the noise (i.e. false positive):
\beq
p=\frac{1}{2|\Delta\widetilde P|^2}\int_{|\widetilde P|^2_{\rm max}}^\infty
\exp\left(-\frac12\frac{|\widetilde P|^2}{|\Delta\widetilde P|^2}\right)d\widetilde P^2=
\exp\left(-\frac12\frac{|\widetilde P|^2_{\rm max}}{|\Delta\widetilde P|^2}\right).
\eeq
This probability was calculated assuming that the real and imaginary components of $\tilde{P}_{\alpha\beta}$ are Gaussian distributed around zero, with standard deviation of $\Delta\tilde{P}_{\alpha\beta}$, which is a consequence of the large $N_\nu(E)$ approximation. Since we probe $n_{\rm tb}$ independent frequencies, the CL is given by the probability that the value at each frequency does not exceed $|\widetilde P|^2_{\rm max}$:
\beq\label{eq:clpmax}
{\rm CL}=\left[1-\exp\left(-\frac12\frac{|\widetilde P|^2_{\rm max}}
{|\Delta\widetilde P|^2}\right)\right]^{n_{\rm tb}}.
\eeq
The $\sigma$-level of a signal is given by
\beq
\sigma{\rm -level}=\sqrt{2}{\rm erf}^{-1}({\rm CL}).
\eeq

Eq. (\ref{eq:clpmax}) provides a simplified form for the Lomb-Scargle (LS) periodogram \cite{Lomb:1976wy,Scargle:1982bw}. While the LS method is more general, our expression has three advantages: It lends itself more easily to interpretation of the result in terms of the parameters of the problem, it does not require the usage of Monte-Carlo simulations, and it allows us to express the CL in terms of our model parameters:
\beq\label{eq:clmphieps}
{\rm CL}=\left[1-\exp\left(-\frac14 \frac{N_\nu^{\rm tot}(E){\rm sinc}^2(m_\phi\Delta t/2)
|\langle\epsilon_{\alpha\beta}(E)\rangle|^2}
{\langle P^0_{\alpha\beta}(E)\rangle(1-\langle P^0_{\alpha\beta}(E)\rangle)}\right)\right]^{\frac{\tau_e}{\Delta t}},
\eeq
where $N_\nu^{\rm tot}(E)=N_\nu(E)n_{\rm tb}$. Thus, given values of $m_\phi$ and $\eta$ (which sets the value of $\epsilon_{\alpha\beta}(E)$), we can calculate the CL at which a given experiment is expected to observe a signal. It might seem from the overall power that increasing $\tau_e$ will counter-intuitively reduce the CL. Doing so however, will also increase $N_\nu^{\rm tot}$, which has a greater impact than the overall power, and will reduce the CL. The sinc term in Eq. (\ref{eq:clmphieps}) may produce some sharp edges in the $m_\phi-\eta$ contours, visible for example in Fig.~\ref{fig:JUNOsensitivity}. These edges are due to aliasing, that occur when $\tau_\phi<\tau_r$. For a point in the parameter space inside these edges, a periodic behaviour will be detected, but with frequency that is smaller than $m_\phi$.

Eq. (\ref{eq:clmphieps}) also suggests the way to optimize the significance of a signal:
\begin{itemize}
\item Increase $\Delta t$ (though that would limit the sensitivity of the experiment to large $m_\phi$ values);
\item Probe energy bins with large $\epsilon_{\alpha\beta}(E)$;
\item Probe energy bins where $\langle P^0_{\alpha\beta}\rangle$ is maximal or minimal;
\item Probe energy bins with large $N_\nu^{\rm tot}$.
\end{itemize}

\subsection*{Acknowledgements}
The authors are grateful to Joachim Kopp for comments on the project, and earlier discussion that seeded this study, and to M.C. Gonzalez-Garcia for fruitful discussions and for helping us to obtain results for KamLAND.
GP also would like to thank Abhishek Banerjee, Nitsan Bar and Hyungjin Kim for discussions.\\
The work of GP is supported by grants from BSF-NSF, Friedrich Wilhelm Bessel research award, GIF, ISF, Minerva, SABRA - Yeda-Sela - WRC Program, the Estate of Emile Mimran, and The Maurice and Vivienne Wohl Endowment. YN is the Amos de-Shalit chair of theoretical physics, and is supported by grants from the Israel Science Foundation (grant number 1124/20), the United States-Israel Binational Science Foundation (BSF), Jerusalem, Israel (grant number 2018257), by the Minerva Foundation (with funding from the Federal Ministry for Education and Research), and by the Yeda-Sela (YeS) Center for Basic Research.

\bibliographystyle{utphys}
\bibliography{ULDMbib}

\end{document}